\documentclass[a4paper,fleqn,usenatbib]{mnras}

\usepackage{newtxtext,newtxmath}
\usepackage[T1]{fontenc}

\DeclareRobustCommand{\VAN}[3]{#2}
\let\VANthebibliography\thebibliography
\def\thebibliography{\DeclareRobustCommand{\VAN}[3]{##3}\VANthebibliography}

\usepackage{graphicx}	% Including figure files
\usepackage{amsmath}	% Advanced maths commands
\title[Instabilities and pulsation in models of MWC\,137]{\huge On the stability and pulsation in models 
of B[e] star MWC\,137}

\author[Parida et al.]{
Sugyan Parida$^{1}$\thanks{E-mail: sugyannitr4142@gmail.com}, Abhay Pratap Yadav$^{1}$\thanks{E-mail: yadavap@nitrkl.ac.in}, 
Michaela Kraus$^{2}$,
Wolfgang Glatzel$^{3}$,
Yogesh Chandra Joshi$^{4}$,
\newauthor \hspace{0.02cm} Santosh Joshi$^{4}$ 
\\
$^{1}$Department of Physics $\&$ Astronomy, National Institute of Technology, Rourkela - 769008, Odisha, India \\
$^{2}$Astronomical Institute, Czech Academy of Sciences, 251 65 Ondřejov, Czech Republic\\
$^{3}$Institut f\"ur Astrophysik und Geophysik, Georg-August-Universit\"at G\"ottingen, Friedrich-Hund-Platz 1, D-37077 G\"ottingen, Germany\\
$^{4}$Aryabhatta Research Institute of Observational Sciences, Manora Peak, Nainital - 263002, India\\ 
}
\date{Accepted XXX. Received YYY; in original form ZZZ}

\pubyear{2023}

\begin{document}
\label{firstpage}
\pagerange{\pageref{firstpage}--\pageref{lastpage}}
\maketitle

\begin{abstract}
B[e] type stars are characterised by strong emission lines, photometric $\&$ spectroscopic variabilities and unsteady mass-loss rates. MWC\,137 is a galactic B[e] type star situated in the constellation Orion. Recent photometric observation of MWC\,137 by TESS has revealed variabilities with a dominant period of 1.9 d. The origin of this variability is not known but suspected to be from stellar pulsation. To understand the nature and origin of this variability, we have constructed three different set of models of MWC\,137 and performed non-adiabatic linear stability analysis. Several low order modes are found to be unstable in which models having mass in the range of 31 to 34 M$_{\odot}$ and 43 to 46 M$_{\odot}$ have period close to 1.9 d. The evolution of instabilities in the non-linear regime for model having solar chemical composition and mass of 45 M$_{\odot}$ leads to finite amplitude pulsation with a period of 1.9 d. 
Therefore in the present study we confirm that this variability in MWC\,137 is due to pulsation. Evolutionary tracks passing through the location of MWC\,137 in the HR diagram indicate that the star is either in post main sequence evolutionary phase or about to enter in this evolutionary phase.
\end{abstract}

% Select between one and six entries from the list of approved keywords.horizontal

%%%%%%%%%%%%%%%%%%%%%%%%%%%%%%%%%%%%%%%%%%%%%%%%%%

%%%%%%%%%%%%%%%%% BODY OF PAPER %%%%%%%%%%%%%%%%%%

\section{Introduction}
Massive stars significantly affect the chemical enrichment and dynamics of galaxies \citep{nomoto_2013, goswami_2022}. 
%The peculiar early-type star 
MWC 137 is a massive galactic B[e] - type star which is surrounded by a optical nebula . \citet{sharpless_1959} first classified this nebula as an H\,II region and named it as Sh 2-266.
For MWC\,137, \citet{ciatti_1975} have found strong H\,I, faint Fe\,II and He\,I lines. 
In addition to that, emissions of O\,I and Ca\,II were also observed in the same study. 
\citet{ciatti_1975} has suggested that the nebula around MWC\,137 is caused by 
the expansion of ejected material from the star. 
%===
\citet{ulrich_1981} has reported the presence of He I recombination 
lines ($\lambda$5876 and $\lambda$10830) in MWC\,137 and linked it with chromospheric emissions. 
This star has been classified as Herbig Ae/Be star by 
 \citet{Hillenbrand_1992}, \citet{berrilli1992infrared} and \citet{the_1994}. 
From the VLA/Australia Telescope survey, \cite{skinner_1993} have reported MWC\,137 as a possible non-thermal
radio source with a significant decrease in radio 
flux on the time scale of $\leq$ 5 months.  
\citet{miro_1994} has taken 19 UBVRIJHK and 12 UBVRI observation for MWC\,137 
using a 1-m telescope at Assy Observatory. 
The author has found variability in all the photometric 
band and classified this star 
as a B0 V type, having a mass - loss of 1.1 $\times$ 10$^{-6}$ M$_{\sun}$ yr$^{-1}$. 
A quasi-periodic variability of 4.07 days is reported for MWC\,137 by \citet{bergner_1994}. 
From narrow band H$_{\alpha}$ image and high resolution spectroscopy of MWC\,137 and its nebula, 
\citet{esteban_1998} concluded that it is not a  Herbig Ae/Be star rather a B[e] type supergiant 
with luminosity L = 2.36 $\times$ 10$^{5}$ L$_{\sun}$. The same authors have suggested that 
the nebula around MWC\,137 (S 266) is likely to 
be produced from the interaction of stellar winds and interstellar medium or
unprocessed ejected material. \citet{henning_1998} have discovered an extended 
bipolar millimeter continuum source in the vicinity of central star MWC\,137. \citet{fuente_2003} 
have reported the presence of circumstellar disk in three B0 stars including MWC 137. These authors have 
derived an upper mass limit of 0.007 M$_{\sun}$   for the disk around this star.  \citet{marston_2006} and 
\citet{marston_2008} have noted the
presence of a diffuse bipolar structure in the ring of nebula around MWC\,137. 
Spectral lines of rings around 
the central star MWC\,137 show that the ring is expanding with velocity of 10-15 km\,s$^{-1}$ indicating 
a dynamical age of 10$^{5}$ yr \citep{esteban_1998}. 
\citet{muratore_2015} have reported that the central star MWC 137 is surrounded by cool and dense ring of CO gas. 
The derived temperature, density and rotational velocity of the ring are horizontal
1900 ($\pm$ 100) K, 3 ($\pm$1) $\times$ 10$^{21}$ cm$^{-1}$ and 84 ($\pm$2) km\,s$^{-1}$, respectively. 
\citet{muratore_2015} have argued that this star is not a Herbig Be star as the gas present in the ring 
is enriched with isotope $^{13}$C. The ratio of $^{12}$C and $^{13}$C has become a diagnostic tool to 
distinguish young, pre-main sequence stars and evolved stars 
\citep[see e.g.,][]{kraus_2009, liermann_2010, muratore_2010a, oksala_2013, kraus_2019}. 
The isotope ratio  $^{12}$C / $^{13}$C  will decrease during the evolution of stars from the 
initial unprocessed interstellar value of 90 to less than 5 \citep{kraus_2019}. The derived value 25 $\pm$ 2 of the 
isotope ratio by \citet{muratore_2015} is appropriate for proto-planetary nebula, main-sequence or
supergiant evolutionary phase.    
%==========
Using VLT MUSE IFU data, \citet{mehner_2016} have discovered a collimated outflow from MWC\,137. The emerging
jet from the star has many knots and the electron density of 10$^{3}$ cm$^{-3}$. 
These authors have concluded that the MWC\,137 is in post main sequence evolutionary phase and 
proposed several possible 
scenarios for the presence of jets and disk including binary system and stellar merger. 
\citet{kraus_2017} have performed multi-wavelength study of this star and its surrounding nebula.  
The authors also have noted that northern part of the nebula is mostly blue shifted while 
the southern part is primarily red shifted. In the same study, it has also been concluded that the 
large scale nebula may not be created and shaped only by MWC\,137. \citet{kraus_2021} have noted a 
substantial variation in the electron density (ranging from 0 to 800 cm$^{-3}$) across the nebula.  The overall 
morphology of the nebula is found to be unchanged within 18 yr. 
%==========
%==========
Using the time series photometric data of Transiting Exoplanet Survey Satellite \citep[TESS,][]{Ricker_2015},
\citet{kraus_2021} have reported a stable variability of 1.93 d for MWC\,137. The cause of this 
variability is unknown. The period of  1.93 d is too short to be originated by rotation or orbital 
motion due to any companion. Therefore it is speculated to be linked with pulsation in MWC\,137. In several 
B-type massive stars, photometric and spectroscopic variabilities have been observed \citep[see e.g.][]{Frost_1902,guthnick_1913} and found to be caused 
by stellar pulsation \citep[e.g.,][]{yadav_2021, yadav_2016}. Since the luminosity to mass ratio
for MWC\,137 is greater than 10$^{3}$ in solar unit therefore dynamical instabilities can be expected in models 
of this star. In models of massive O and B type stars, it has been found that these instabilities 
may lead to pulsation, surface eruptions and re-arrangement 
of stellar structure \citep[see e.g.,][]{glatzel_1999, yadav_2016, yadav_2017b}. 
In the present study, we intend to investigate instabilities and their 
consequences in models of MWC\,137. Instabilities identified in the linear stability analysis will be further 
considered for nonlinear numerical simulations to find out the final fate of unstable models. Comparing the 
obtained numerical results with observed variabilities in MWC\,137 will allow us to confirm or rule out pulsation 
as the cause of reported 1.9 d variability. 
%==========
Photometric variability of MWC\,137 using the data of TESS is presented in section \ref{pv}.
In section \ref{models}, properties of considered models of MWC\,137 are described. 
Linear stability analysis and associated results are presented in section \ref{lsa}. 
Nonlinear numerical simulation for selected unstable models is mentioned in section \ref{nlr}. Evolutionary status of this star is highlighted in section \ref{es}. Discussion is mentioned in section \ref{DC} followed by conclusions in section \ref{conclu}.

\section{Photometric variability of MWC 137 from TESS} \label{pv}
\begin{figure*}
\centering $
\Large
\begin{array}{cc}

   \scalebox{0.69}{ \input{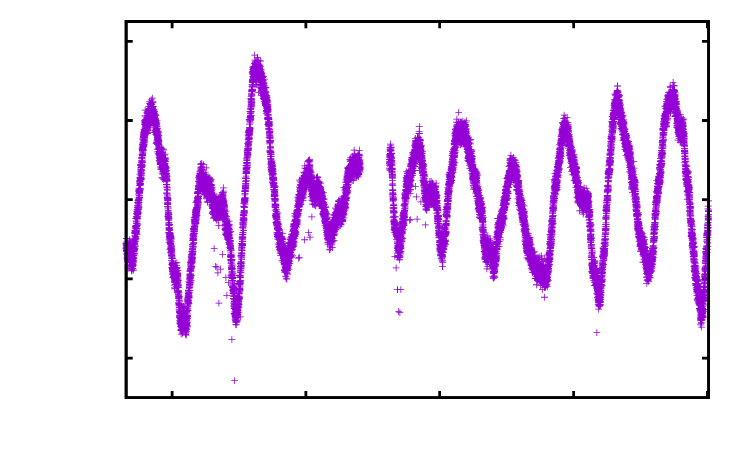} } 
   \scalebox{0.69}{ \input{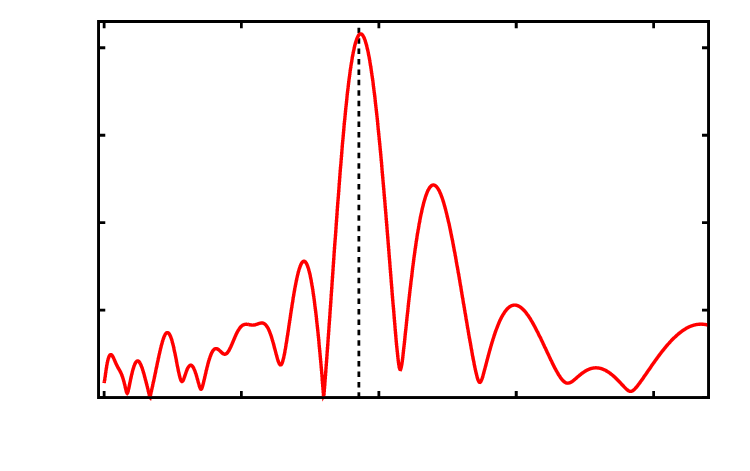} } \\
 \end{array}$
 \caption{Lightcurve (left) and Periodogram (right) of MWC\,137 observed under sector 6 by TESS. Variability in the lightcurve and a dominant period 1.9 d are present.}
 \normalsize
 \label{p1}
 \end{figure*}

\begin{figure*}
\centering $
\Large
\begin{array}{cc}

   \scalebox{0.69}{ \input{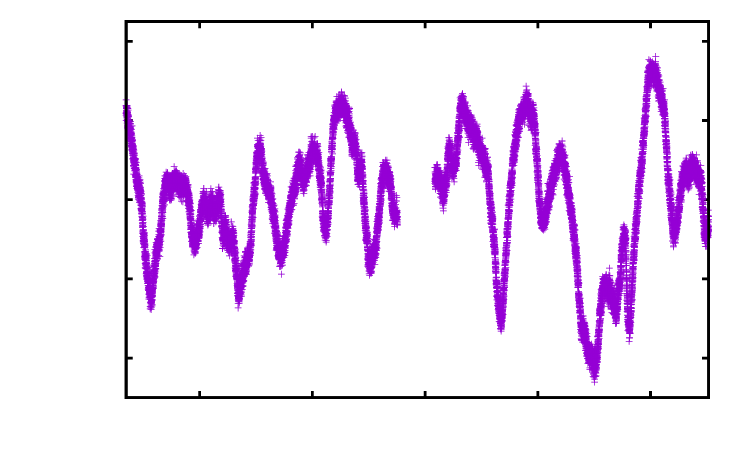} } 
   \scalebox{0.69}{ \input{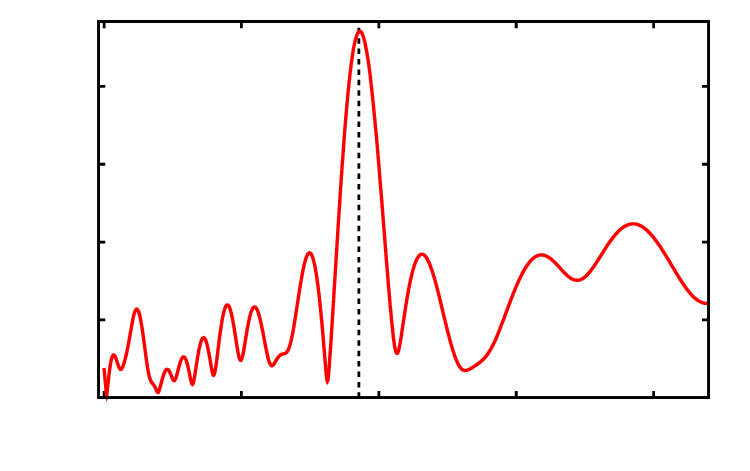} } \\
 \end{array}$
 \caption{Lightcurve (left) and Periodogram (right) of MWC\,137 observed under sector 33 by TESS. Variability in the lightcurve and a dominant period 1.9 d are present.}
 \normalsize
 \label{p2}
 \end{figure*}

The normalized light curves for MWC\,137 were generated using data from sectors 6 and 33 of TESS. The data for both sector 6 and sector 33 was taken in a 2 yr gap with high cadence (120 s) and are displayed in Fig.\ref{p1} and Fig.\ref{p2}, respectively.
Further in order to study the TESS time series in the frequency domain, periodograms / Fourier analysis is required. For TESS observations, we have computed Lomb-Scargle periodogram as they are more appropriate for unevenly spaced data.
This Lomb-Scargle analysis is done using the Lightkurve package \citep{2018ascl.soft12013L}.
For each sector, we have detected a dominant period of 1.9 d, as is evident from Fig. \ref{p1}, \ref{p2}. The same variability is also reported in a recent paper by \citet{kraus_2021} using PERIOD04 \citep{periodo4}.

\section{Models for MWC\,137} 

\label{models}
%==========
It is challenging to determine fundamental parameters such as mass, 
luminosity and surface temperature for B[e] type massive stars. To construct the models of MWC\,137, we have adopted the 
average value of surface
temperature T$_{\rm{eff}}$ = 28200 K and luminosity log L/L$_{\sun}$ = 5.84 from \citet{kraus_2021} . 
There is a large uncertainty in the distance of MWC\,137 which ranges from 1 kpc to 13 kpc. \citet{kraus_2021} have derived 
the value of luminosity using the distance 5.15 kpc. This distance is in agreement with the value (5.2 kpc)
determined by \citet{mehner_2016}. 
There is a large uncertainty in the mass of MWC\,137 hence models having a sequence of mass in the range of 
27 to 70 M$_{\sun}$ have been considered. This mass range 
includes the mass 37$^{+9}_{-5}$ M$_{\sun}$ derived by \citet{kraus_2021} for this star. The present study is focused on the models of this star having solar chemical composition (Z = 0.02, Y = 0.28, X = 0.70). Additionally, to study the dependency of instabilities on metals, we have also considered models having Z = 0.01 and Z = 0.03.  

For constructing the models of MWC\,137, we have integrated stellar structure equations --  mass conservation, 
momentum conservation, energy conservation and energy transport -- from the surface upto 
a point where temperature is 10$^7$ K. With this integration, we obtain envelope models of the star to 
perform linear stability analysis followed by nonlinear simulations. Since instabilities are mostly found in the 
envelopes of massive O and B type stars therefore we have restricted our present study in envelopes to 
avoid the complications of nuclear reactions in the core. 
Schwarzschild criterion is applied for convection. Standard mixing length theory is used with the mixing 
length parameter $\alpha$ = 1.5 for convection \citep{bohm_1958}. 
Rotation and magnetic fields have been disregarded. 
For the opacity, OPAL tables \citep{rogers_1992, rogers_1996, iglesias_1996} are used.

\section{Stability analysis for models of MWC\,137}  \label{lsa}
Linearised pulsation equations to perform the stability analysis for radial perturbation have been adopted from \citet{gautschy_1990a}. This set of equations form a fourth order boundary eigenvalue problem. Out of four boundary conditions, two boundary condition are applied at the surface and remaining two are used at the bottom of the envelope. The eigenvalue problem is solved using the Ricatti method as described by \citet{gautschy_1990a}. Obtained eigenfrequencies are complex where real part denotes pulsation period and imaginary part indicates damping ($\sigma_i$ > 0) or excitation ($\sigma_i$ < 0). The calculated eigenfrequencies are normalised by global free fall timescale.

\begin{figure*}
\centering $
\Large
\begin{array}{cc}

   \scalebox{0.69}{ \input{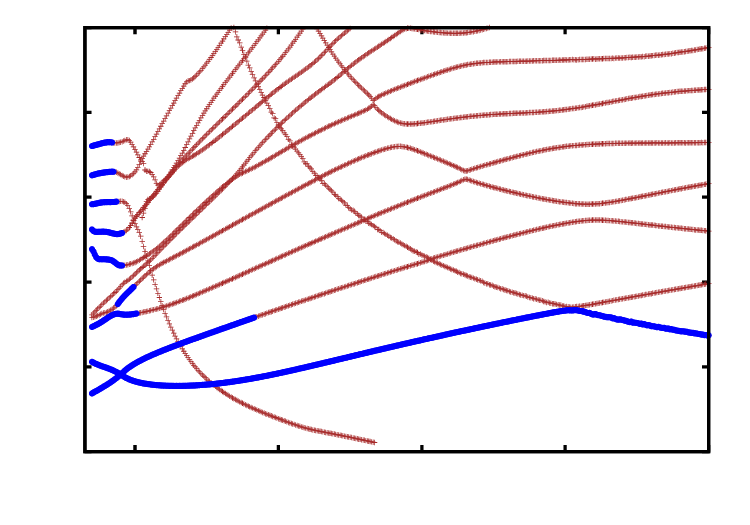} } 
   \scalebox{0.69}{ \input{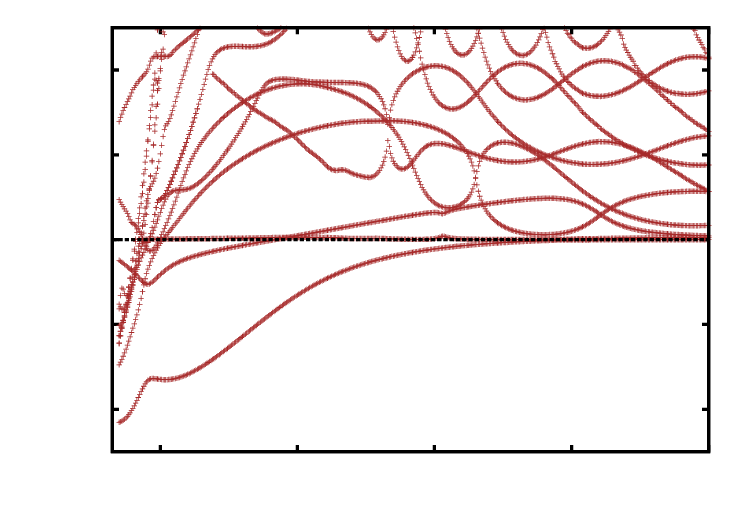} } \\
 \end{array}$
 \caption{Real (left) and imaginary (right) parts of eigenfrequencies are plotted as a function of mass for models of MWC\,137 in the mass range between 27 to 70 M$_{\sun}$ having solar chemical composition (Z = 0.02). These eigenfrequencies are normalised by the global free fall timescale. Negative imaginary part and blue lines in real part of eigenfrequencies corresponds to unstable (excited) modes. For these models, value of log T$_{\rm{eff}}$ (K) = 4.45 and luminosity log L/L$_{\sun}$ = 5.84 have been used. } 
 \normalsize
 \label{modal_1}
 \end{figure*} 

\begin{figure*}
\centering $
\Large
\begin{array}{cc}

   \scalebox{0.69}{ \input{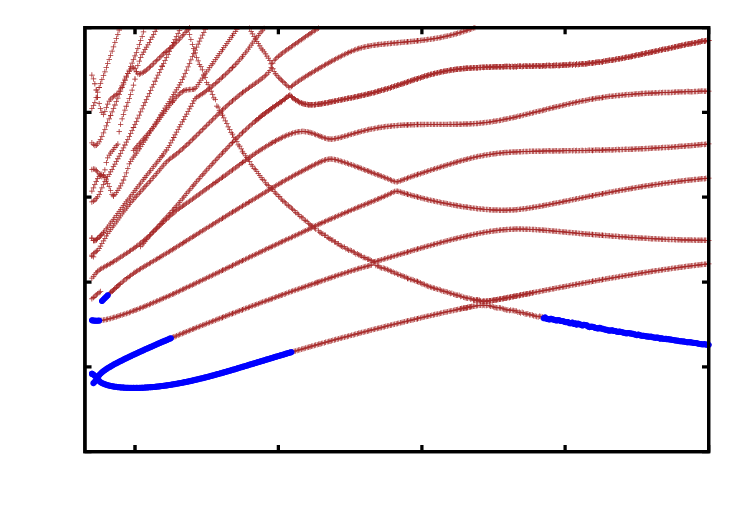} } 
   \scalebox{0.69}{ \input{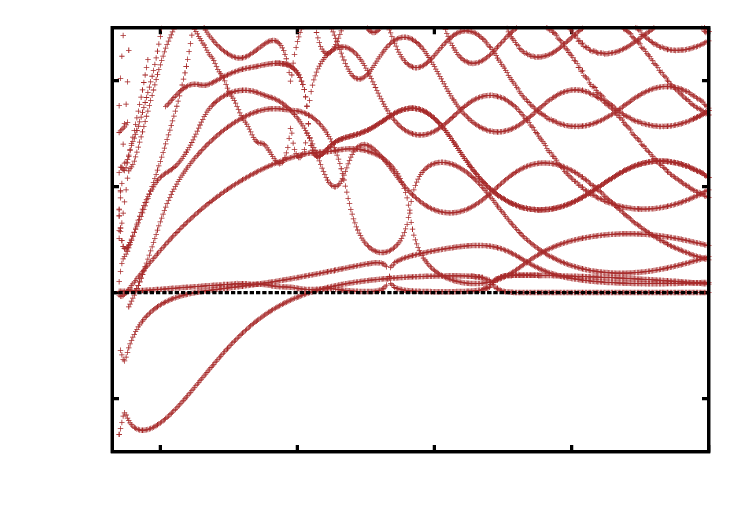} } \\
 \end{array}$
 \caption{Same as Fig. \ref{modal_1} but for Z = 0.01 }
 \normalsize
 \label{modal_2}
 \end{figure*}

 \begin{figure*}
\centering $
\Large
\begin{array}{cc}

   \scalebox{0.69}{ \input{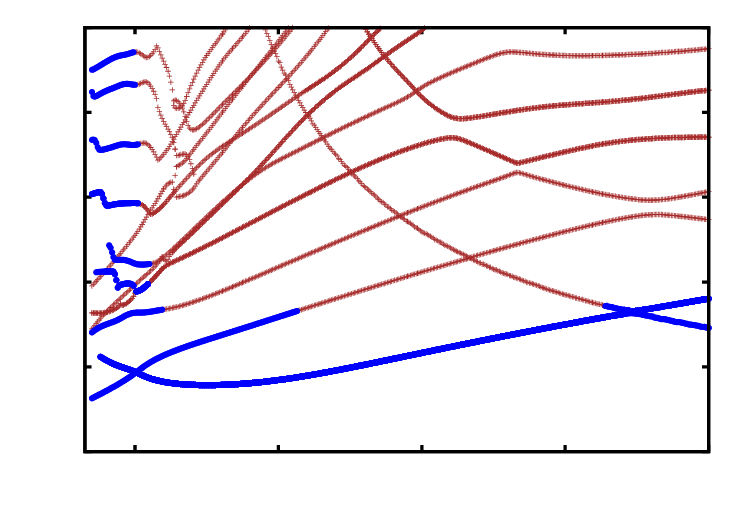} } 
   \scalebox{0.69}{ \input{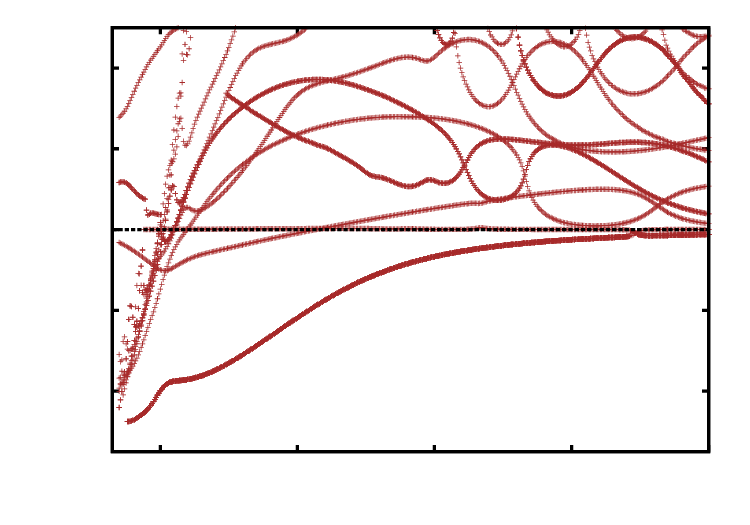} } \\
 \end{array}$
 \caption{Same as Fig. \ref{modal_1} but for Z = 0.03}
 \normalsize
 \label{modal_3}
 \end{figure*} 

\begin{figure*}
\centering $
\Large
\begin{array}{cc}

   \scalebox{0.69}{ \input{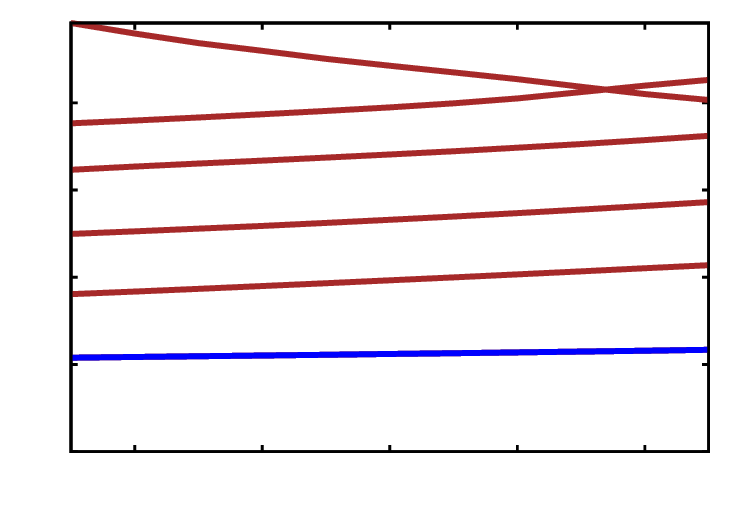} } 
   \scalebox{0.69}{ \input{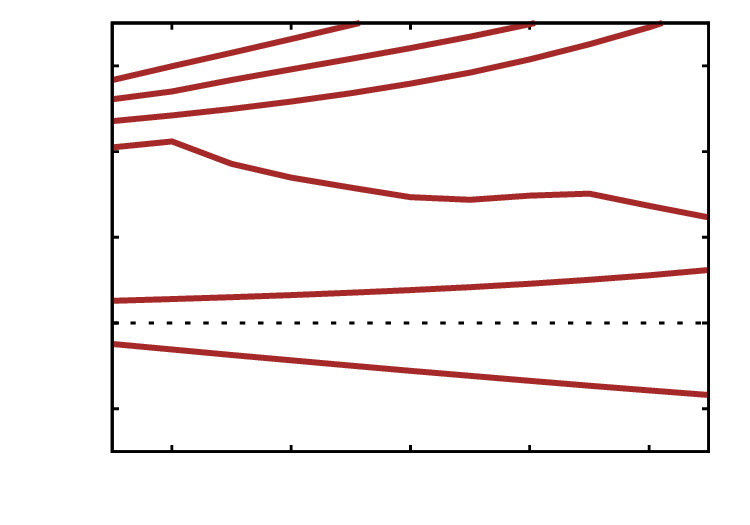} } \\
 \end{array}$
 \caption{Same as Fig. \ref{modal_1} but here eigenfrequencies are plotted as a function of surface temperature. \citet{kraus_2021} has suggested that the surface temperature of the MWC\,137 is in the range of 4.40 $<$ log T$_{\rm{eff}}$ $<$ 4.49. For these models, mass and luminosity (log L/L$_{\sun}$) are 45 M$_{\odot}$ and 5.84, respectively. }
 \normalsize
 \label{modal_4}
 \end{figure*} 

\begin{figure*}
\centering $
\Large
\begin{array}{cc}

   \scalebox{0.69}{ \input{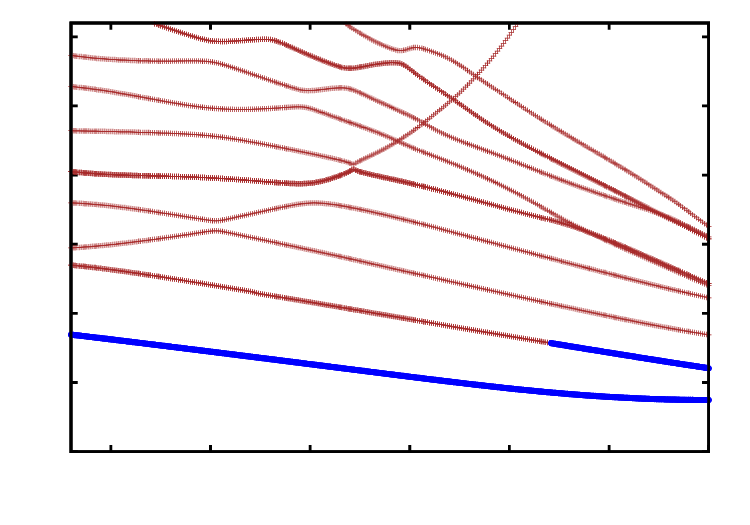} } 
   \scalebox{0.69}{ \input{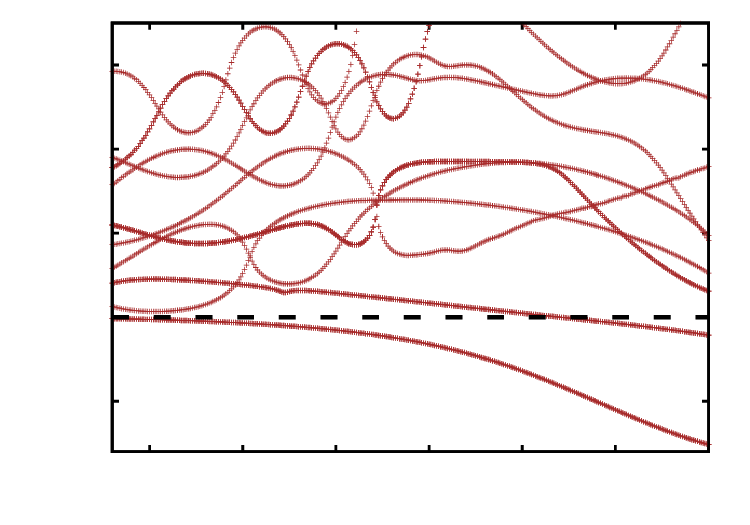} } \\
 \end{array}$
 \caption{Same as Fig. \ref{modal_1} but here eigenfrequencies are plotted as a function of luminosity. \citet{kraus_2021} has suggested that the luminosity of the MWC\,137 is in the range of 5.71 $<$ log L/L$_{\sun}$ $<$ 5.97. For these models, mass and surface temperature (log T$_{\rm{eff}}$) are 45 M$_{\odot}$ and 4.45, respectively. }
 \normalsize
 \label{modal_5}
 \end{figure*}

The outcome of linear stability analysis is generally depicted in a form of `modal diagram'. In the modal diagram, real and imaginary part of eigenfrequencies are given as a function of fundamental stellar parameter such as mass or surface temperature. Modal diagram for the case of models having solar chemical composition is given in Fig. \ref{modal_1}, where real part of the eigenfrequency is mentioned in Fig. \ref{modal_1}(left) and imaginary part is given in \ref{modal_1}(right). Thick blue lines in Fig. \ref{modal_1}(left) correspond to unstable modes where imaginary part of eigenfrequency is negative. All of the considered sequence of models starting from 27 M$_{\odot}$ to 70 M$_{\odot}$ are unstable. Number of unstable modes is increasing for models having higher luminosity to mass ratio. We also note mode interaction phenomena which is frequently present in the models of high luminosity to mass ratio. One of the modes is unstable for all the considered models from 27 M$_{\odot}$ to 70 M$_{\odot}$. The strength of instability associated with this mode increases for lower mass models. In the vicinity of this mode, another mode is unstable for models having mass below 40 M$_{\odot}$. For the model of mass 29 M$_{\odot}$, two low order unstable modes are showing interaction in real part of eigenfrequeny while their imaginary parts are well separated. Real part of eigenfrequency associated with one of the damped modes is approaching towards zero for model having mass 48 M$_{\odot}$.

Linear stability analysis has also been performed for models having  following chemical compositions: 

\begin{enumerate}
    \item X = 0.71, Y = 0.28, Z = 0.01
    \item X = 0.69, Y = 0.28, Z = 0.03
\end{enumerate}
The modal diagram for models having Z = 0.01 is given in Fig. \ref{modal_2}. In this case, models only in the mass range between 27 M$_{\odot}$ to 41 M$_{\odot}$ and 59 M$_{\odot}$ to 70 M$_{\odot}$ are unstable. Number of unstable modes and strength of instabilities have reduced compared to models having Z = 0.02. All the considered higher order modes are damped. 

For the models having Z = 0.03, modal diagram is given in Fig. \ref{modal_3}. In this case, all the considered models in the mass range 27 M$_{\odot}$ to 70 M$_{\odot}$ are unstable. The modal diagram for Z = 0.03 is qualitatively similar to the Z = 0.02. However, for the case of Z = 0.03, one additional mode is unstable for models having mass above 62 M$_{\odot}$. Strength of instabilities associated with the unstable modes are slightly more than the models with Z = 0.02.    

%NEW ADDITION =============
Outcomes of the linear stability analyses mentioned in the modal diagrams (Fig. \ref{modal_1} to Fig. \ref{modal_3}) are based on the models of MWC\,137 having surface temperature of 28200 K (log  T$_{\rm{eff}}$ $\approx$ 4.45). However, the previous studies have shown that the value of log T$_{\rm{eff}}$ for this star is in the range of 4.40 to 4.49 \citep{kraus_2021}. To study the presence and strength of instabilities as a function of surface temperature, we have also carried out linear stability analysis for the models having mass of 45 M$_{\odot}$ with solar chemical composition and surface temperature (log T$_{\rm{eff}}$) in the range of 4.40 to 4.49. The result is presented in the Fig. \ref{modal_4} where real and imaginary part of eigenfrequencies are plotted as a function of surface temperature. Similar to the modal diagram of Fig. \ref{modal_1}, blue lines in the real part and negative imaginary part indicate unstable modes. From this figure, we infer that all the considered models of MWC 137 in the surface temperature (log Teff) range of 4.40 to 4.49 are unstable and pulsation period associated with unstable mode of different models is almost same.

Similar to the ambiguity in the determination of surface temperature, luminosity of the star is also subject to possible errors mostly due to distance determination. \citet{kraus_2021} have determined the luminosity log L/L$_{\sun}$ = 5.84 with an error of ± 0.13 for this star. Earlier studies have shown that the high luminosity to mass ratio can induce several dynamical instabilities such as strange mode instabilities \citep{glatzel_1990NAR,kiriakidis1993stability}. To examine the luminosity dependency in the present case we have selected models with solar chemical composition having mass of 45 M$_{\odot}$, log T$_{\rm{eff}}$ = 4.45 and luminosity in the range of log L/L$_{\sun}$ = 5.68 to log L/L$_{\sun}$ = 6.00. The outcomes are given in the modal diagram mentioned in Fig. \ref{modal_5} where real and imaginary parts of the eigenfrequencies are plotted as a function of luminosity. In total, two modes are unstable in the considered luminosity range. One of the low order modes is unstable in all the models while the other mode is only unstable in models having log L/L$_{\sun}$ > 5.92. For both unstable modes, strength of instabilities is increasing in models having higher luminosity to mass ratios which is expected and in agreement with previous studies \citep[see e.g.][]{glatzel_1994, saio_1998}.

\begin{figure}
\centering $
\Large
\begin{array}{c}
 \scalebox{0.62}{ \input{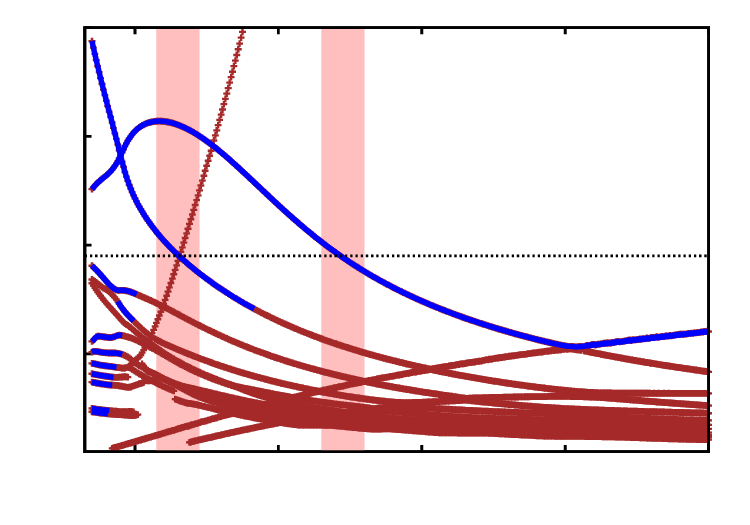} } \\
 \end{array}$
 \caption{Period associated with different modes are plotted as a function of mass for models of MWC\,137 having solar chemical composition (z = 0.02). Unstable modes are represented by blue lines. Horizontal dotted line is indicating the dominant period of observed variability. Vertical brown columns represent the location of models having period close and equal to 1.9 d. Same as Fig. \ref{modal_1}, value of log T$_{\rm{eff}}$ (K) = 4.45 and luminosity log L/L$_{\sun}$ = 5.84 have been used for these models. } 
 \normalsize
 \label{periods}
 \end{figure}

For models having solar chemical composition (Z = 0.02), periods associated with different modes as a function of mass are given in Fig. \ref{periods}. In this figure, thick blue lines indicate unstable modes. A dotted horizontal line represents the observed period of 1.9 d. The vertical pink columns in Fig. \ref{periods}, represent two different mass range (31 to 34 M$_{\odot}$ and 43 to 46 M$_{\odot}$) where unstable modes have period close to 1.9 d. Periods associated with two unstable modes for models having mass close to 33 M$_{\odot}$ and 45 M$_{\odot}$ are exactly matching with the observed period of 1.9 d. Model having mass of 33 M$_{\odot}$ has two unstable modes with a period of 1.9 d and 3.1 d. Strength of instability associated with the mode having period 3.1 d is more than the mode with the period of 1.9 d (see Fig. \ref{modal_1}). Therefore unstable mode with the period of 3.1 d is likely to dominate over the other unstable mode. Periods of all the unstable models are below 4 d in the considered mass range.

% ================ Section 3 ===============
% ================ Section 3 ===============

%\section{Stability analysis}

\begin{figure*}
\centering $
\huge
\begin{array}{cccccc}
  \scalebox{0.455}{ \input{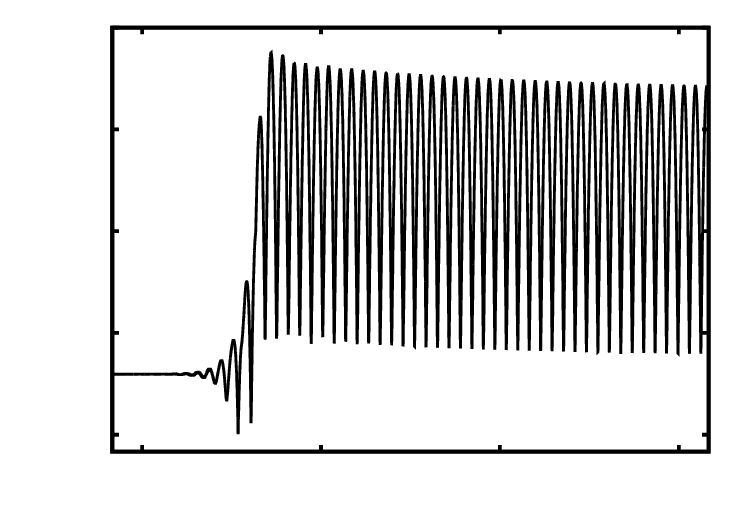}}  
   \scalebox{0.455}{ \input{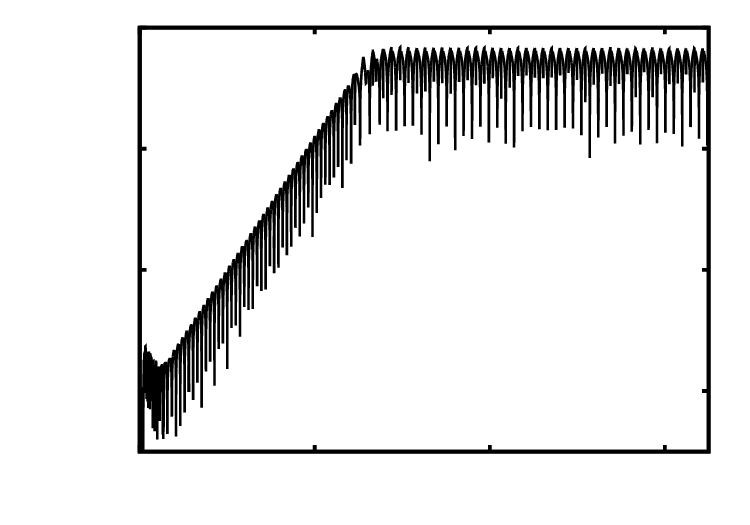} }
  \scalebox{0.455}{ \input{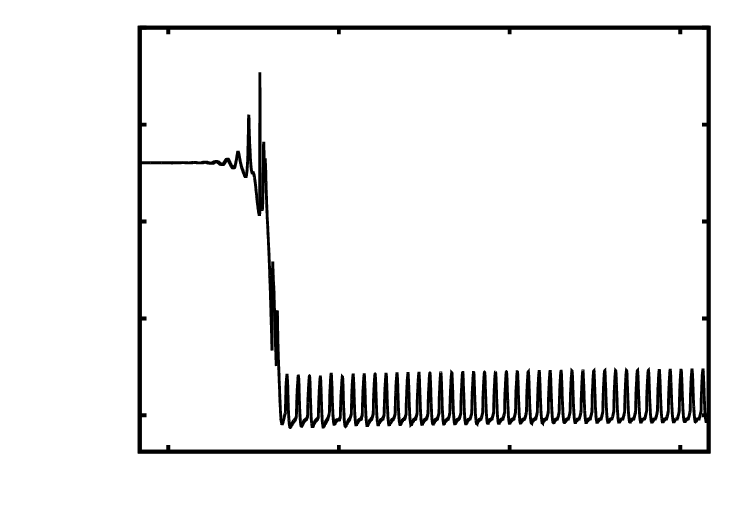} } \\
  \scalebox{0.455}{ \input{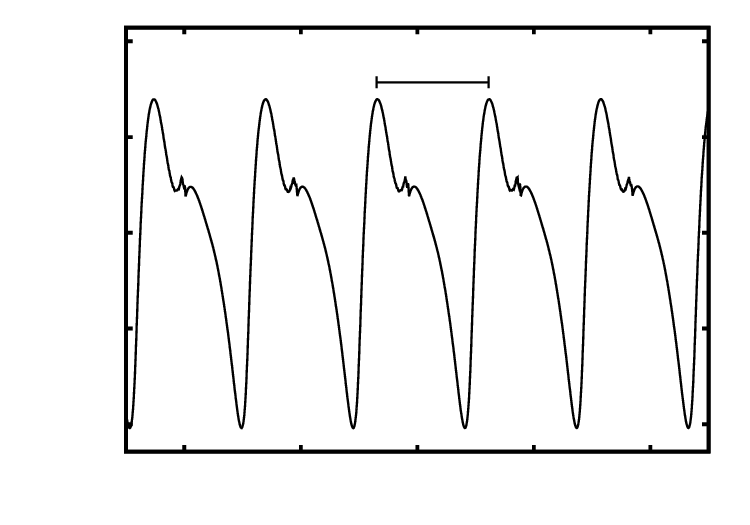} } 
   \scalebox{0.455}{ \input{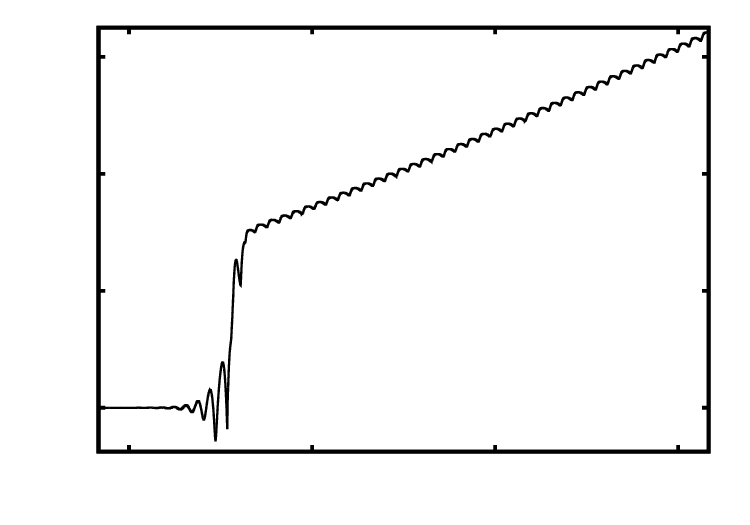} }
  \scalebox{0.455}{ \input{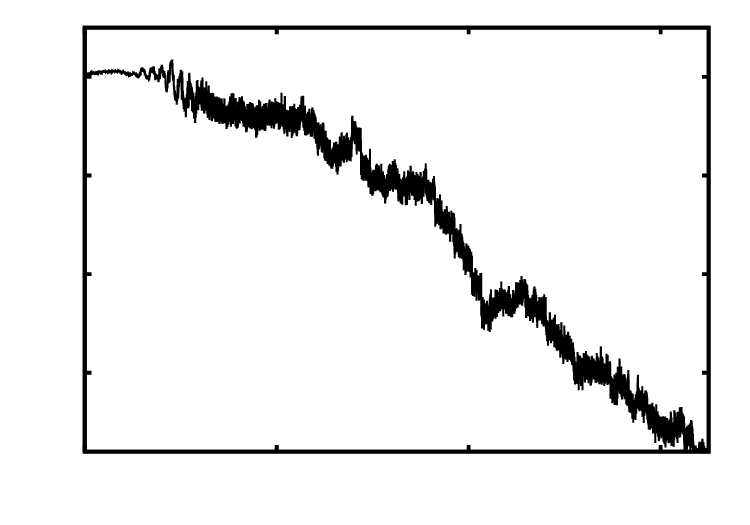} } \\
   \end{array}$
 \caption{Evolution of instability in a model of MWC\,137 having mass of 45 M$_{\odot}$ and solar chemical composition: Variation in radius (a), variation in velocity (b) and temperature (c) associated with outermost grid point are plotted as a function of time. Changes in the bolometric magnitude (d), time integrated acoustic energy (e) and error in the energy balance (f) are given as a function of time. Instabilities lead to finite amplitude pulsation with a period of 1.9 d which is clearly visible in the profile of bolometric magnitude.  }
 \normalsize 
 \label{45m_nonlin}
 \end{figure*}

\begin{figure*}
\centering $
\huge
\begin{array}{ccc}
  \scalebox{0.455}{ \input{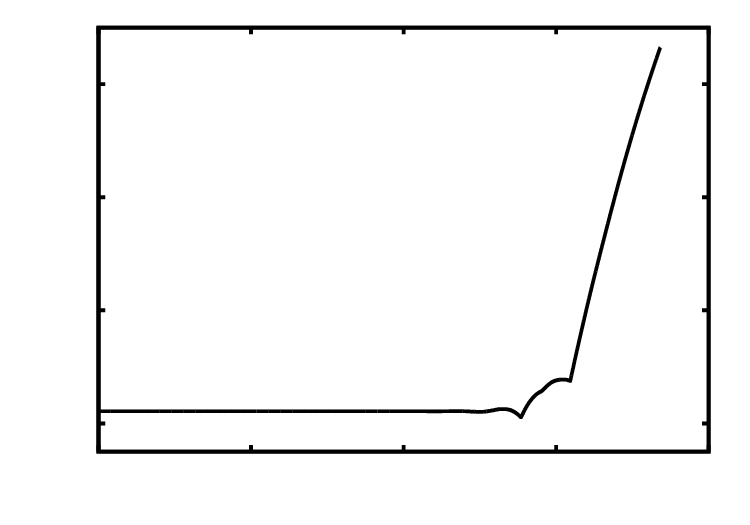} }
   \scalebox{0.455}{ \input{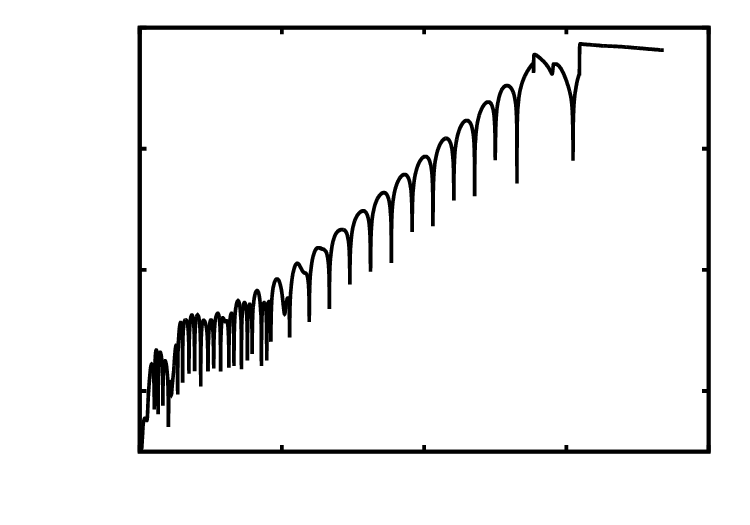} }
  \scalebox{0.455}{ \input{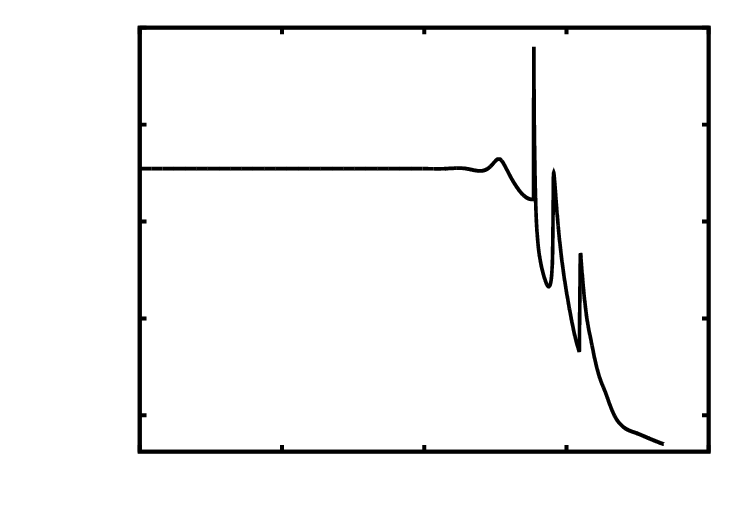} } \\
   \end{array}$
 \caption{Evolution of instability in nonlinear regime for a model of MWC 137 having mass of 33 M$_{\odot}$.}
 \normalsize
 \label{33m_nonlin}
 \end{figure*}

\section{Instabilities in non-linear regime}  \label{nlr}
Linear stability analysis has revealed that several 
low order modes are unstable in the considered three sets of models for MWC 137. To determine the final fate of unstable models, we have followed instabilities in nonlinear regime using the similar set of equations as described by \citet{grott_2005} for selected models of MWC 137. The adopted numerical scheme is energy conservative which is essential for the reliability of simulation's outcomes. To handle the shock waves, artificial viscosity is introduced \citep[see e.g.][]{grott_2005} with the value of parameter 10. Sensitivity of the numerical outcomes on the artificial viscosity parameters are discussed in detail by \citet{grott_2005} and \citet{yadav_2018}. During the non-linear simulation of O and B type stars, the used value (10) of artificial viscosity is found to be adequate for smearing out discontinuities and suppressing Gibbs's phenomena \citep{yadav_2016, yadav_2018}. For nonlinear numerical simulations, four boundary conditions are used. Two of them are imposed at the bottom of the envelope for which zero velocity and a constant flux are considered. The remaining two are applied at the surface where the gradient of compression is set to zero and no heat is stored. This choice of outer boundary conditions will enable the shock waves to pass without being reflected \citep[for a detailed discussion the reader is referred to][]{grott_2005}.

Outcome of nonlinear simulation for two of the considered models having mass of 45 M$_{\odot}$ and 33 M$_{\odot}$ are given in Fig. \ref{45m_nonlin} and \ref{33m_nonlin}, respectively. The outcome of these two models are presented here as these models show unstable mode in linear stability analysis with the period of 1.9 d. In Fig. \ref{45m_nonlin}, variation in the radius \ref{45m_nonlin}(a), velocity \ref{45m_nonlin}(b) and temperature \ref{45m_nonlin}(c) associated with outermost grid point, change in the bolometric magnitude \ref{45m_nonlin}(d), time integrated acoustic energy \ref{45m_nonlin}(e) and error in the energy balance \ref{45m_nonlin}(f) are given as a function of time. Evolution of instabilities starts from an initial value 10$^{-4}$ cm/s for velocity. Code picks up the instability from this value of numerical noise without external perturbation and enter into the phase of exponential growth. After 50 days velocity amplitude saturates with a value close to 157 km/s (Fig. \ref{45m_nonlin}b). The final value of velocity amplitude is approximately 22 percent of the escape velocity of this model. After entering into the non-linear regime, the mean radius of the model has increased to 2.8 $\times$ 10$^{12}$ cm from the initial hydrostatic value of 2.3 $\times$ 10$^{12}$ cm (Fig. \ref{45m_nonlin}a). Due to this inflation of the radius, a drop in surface temperature is expected. However, due to ambiguity in the outer boundary condition, the temperature associated with the outermost grid point may not necessarily represent surface temperature of the star (Fig. \ref{45m_nonlin}c). Variation in the bolometric magnitude is clearly showing a period of 1.9 d (Fig. \ref{45m_nonlin}d). This 1.9 d period is also present in the variation profile of radius, velocity and temperature. The value of time integrated acoustic energy is of the order of 10$^{40}$ erg (Fig. \ref{45m_nonlin}e) which is four order of magnitude larger than the error in the energy balance. This time integrated acoustic energy (4$\pi$R$^{2}$vp, where p and v stand for pressure and velocity perturbations at the model's outer boundary, and R is the radius of that boundary) is representing the mechanical energy lost from the system in the form of acoustic waves. As described by \citet{yadav_2017b}, during a pulsation period, phases of incoming and outgoing fluxes are present. While integrating for a complete cycle, the outgoing energy is slightly higher than the incoming energy. Resultantly, the integrated acoustic energy is increasing as a function of time after the onset of the regular pulsation cycle in the simulation. The slope of the integrated acoustic energy can be used to determine the mean mechanical luminosity \citep[see also,][]{grott_2005}.
 In the performed non-linear numerical simulation, the error in the energy balance is several order of magnitude smaller than time integrated acoustic energy and kinetic energies. It represents the energy conservation which is essential in the modelling of stellar instabilities and pulsation \citep[see e.g.][]{grott_2005, yadav_2018}. Similar to the model of 45 M$_{\odot}$, evolution of radius, variations associated with the outermost grid point of velocity and temperature for a model having mass of 33 M$_{\odot}$ are given in Fig. \ref{33m_nonlin}. In the nonlinear regime, we note a substantial inflation as the radius becomes more than seven times the initial hydrostatic radius (Fig. \ref{33m_nonlin}a). Velocity variations (Fig. \ref{33m_nonlin}b) show that the code picks-up the instability from the numerical noise and it goes through the phase of exponential growth. Due to the inflation, as noted in the radius, temperature associated with the outermost grid is dropping and attaining a value less than 7000 K (Fig. \ref{33m_nonlin}c). Due to unavailability of opacity table for further calculations, numerical simulation had to be terminated. Linear stability analysis has shown that this model has two unstable modes with periods of 1.9 d and 3.1 d. During the phase of oscillatory exponential growth, a period of 3.1 d is present which corresponds to the more strongly unstable mode.

\begin{figure*}
\centering $
\huge
\begin{array}{cccc}
  \scalebox{0.48}{ \input{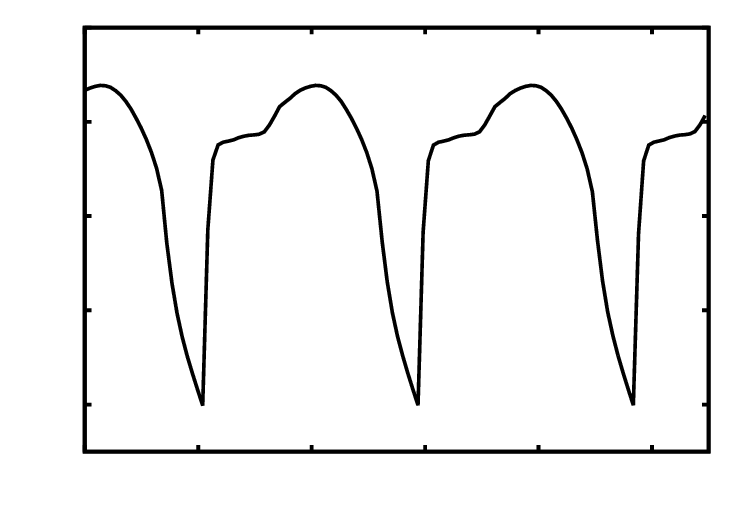} }
   \scalebox{0.48}{ \input{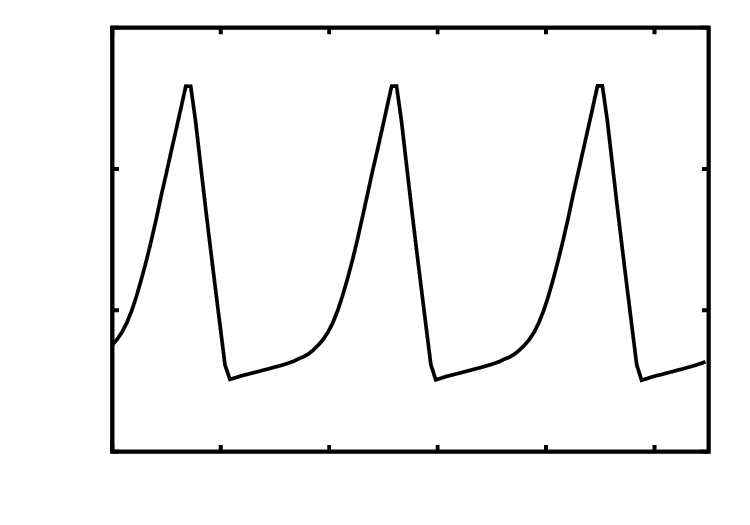} } \\
  \scalebox{0.48}{ \input{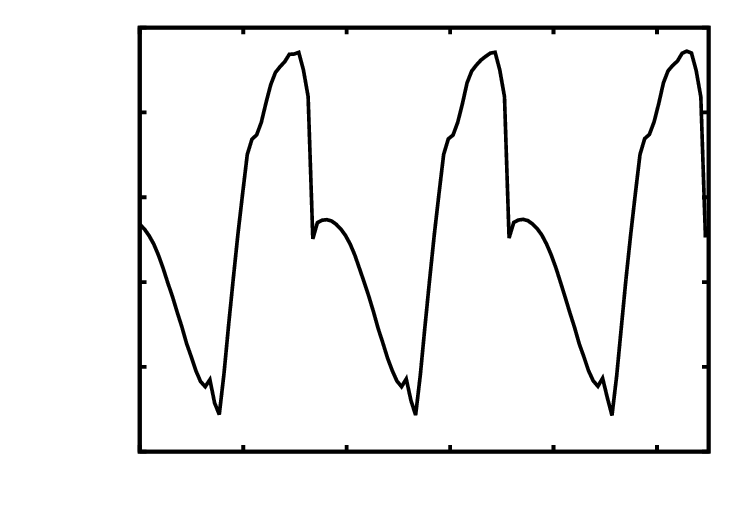} }
  \scalebox{0.48}{ \input{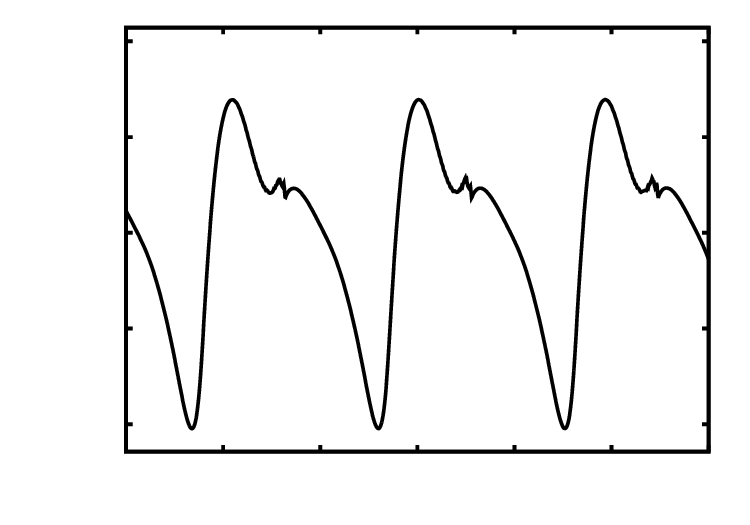} } \\
\end{array}$
  \caption{Variations in velocities (a), radius (b), temperature (c) associated with the grid point close to the photosphere and bolometric magnitude (d) of the outermost gridpoint as a function of normalised time for the model with mass of 45 M$_{\odot}$.}
 \normalsize
 \label{45mphot}
 \end{figure*}

As pointed out earlier, outermost grid point used for the numerical simulation  may not necessarily represents the photosphere of the star. Therefore for direct comparison with observation, variation of quantities in the vicinity of photosphere is more useful. 
For a model having mass of 45 M$_{\odot}$, variation in velocity, radius and temperature associated with the grid point in the vicinity of photosphere as a function of normalised time is given in Fig. \ref{45mphot} for three pulsation cycles. In addition to these quantities, variation in the bolometric magnitude of the outermost grid point is also mentioned in Fig. \ref{45mphot}.
For this purpose, the grid point is selected where $\sigma$T$^4$ (where $\sigma$ is Stefan-Boltzmann constant and T represents temperature) is acquiring an equivalent value of the total flux. The location of the grid point satisfying the adopted criteria is changing during the pulsation cycles. Period of 1.9 d is also present at this grid point in the variation profile of velocity, radius and temperature (see Fig. \ref{45mphot}). Temperature and radius variations are in the range of 25300 K to 29800 K and 2.35 $\times$ 10$^{12}$ cm to 2.95 $\times$ 10$^{12}$ cm, respectively. Variation in the bolometric magnitude is approximately in the range of -0.15 to +0.10 during these three cycles. 
The shape of the magnitude variation profile of Fig. \ref{45m_nonlin} and Fig. \ref{45mphot} is the same as these variations are connected with the outermost grid point. The pulsation period of 1.9 d is present in both the cases.

\section{Evolutionary state of MWC\,137} \label{es}
\begin{figure}
\centering$
\huge
\begin{array}{c}
  \scalebox{0.44}{ \input{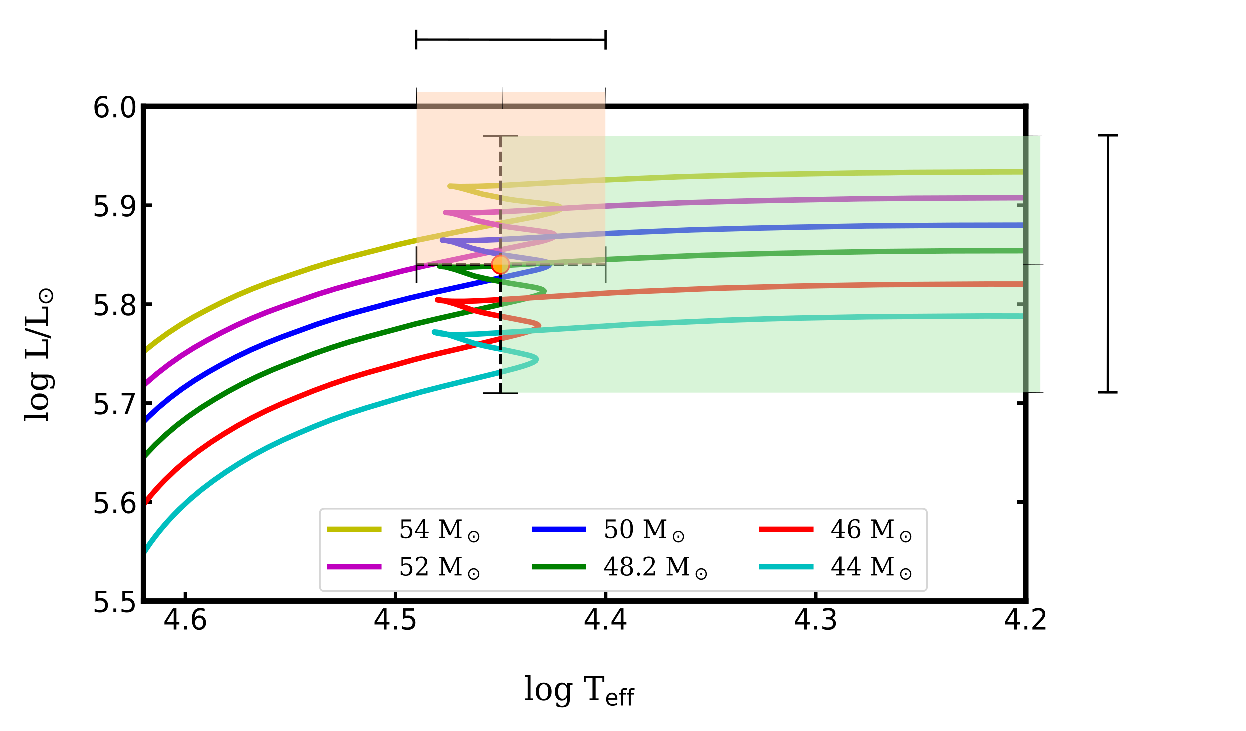} } \\
\end{array}$
 \caption{Evolutionary tracks of stellar models having solar chemical composition. The position of MWC\,137 including errors in the surface temperature and luminosity is indicated by orange dot. Initial masses associated with the evolutionary tracks are given with the corresponding color code. Periods associated with the most unstable mode for the models having mass of 45 M$_{\odot}$ are also mentioned for the range of errors in surface temperature and luminosity.}
 \normalsize
 \label{fig:evol}
 \end{figure}

Linear stability analysis and non-linear numerical simulation have shown that 1.9 d variability can be explained by stellar pulsation as found in the models having mass of approximately 45 M$_{\odot}$. Evolutionary stage of many B[e] type stars including MWC\,137 is uncertain \citep{mehner_2016,kraus_2019,kraus_2021}. To determine the evolutionary state of MWC\,137, we have plotted evolutionary tracks with initial masses of 44  M$_{\odot}$, 46 M$_{\odot}$, 48.2 M$_{\odot}$, 50 M$_{\odot}$, 52 M$_{\odot}$ and 54  M$_{\odot}$
in Fig. \ref{fig:evol} using solar chemical composition with the help of MADSTAR \footnote[1]{\url{http://user.astro.wisc.edu/~townsend/static.php?ref=ez-web}}
 which is based on the stellar evolution code Evolve ZAMS \citep{paxton_2004}. The Evolve ZAMS (EZ) can evolve the stellar models having masses in the range of 0.1 to 100 M$_{\odot}$ with several possible values of metallicity. It creates one-dimensional, non-explosive, non-relativistic and non-rotating stellar models using OPAL opacity tables. For more details on the adopted numerical scheme and parameters used for the stellar evolution, we refer to \citet{paxton_2004} and references therein.   
 The evolutionary track of 48.2 M$_{\odot}$  is passing through the location of MWC\,137 after the over all contraction phase. The location of MWC\,137 is marked using a red circle with the errors in the luminosity and surface temperature in Fig. \ref{fig:evol}. The location of this circle is fixed using the average value of luminosity and surface temperature of this star as suggested by \citet{kraus_2021}. Due to ambiguity in the distance of this star, error in the luminosity is relatively high and consequently leading to uncertainty in the mass determination using evolutionary tracks. Considering the location of MWC\,137 including the error bars in the Hertzsprung-Russell (HR) diagram and evolutionary tracks for different massive stars, it is evident that MWC\,137 is either a post main sequence star or about to enter in the post main sequence evolution. We infer from the evolutionary track of the model having initial mass 48.2  M$_{\odot}$ that the star has just crossed the overall contraction phase and currently H-buring is taking place in shell around the He-core. 
 The derived age for the star from this evolutionary track is 3.9 $\times$ 10$^{6}$ yr which is in the range as estimated by \citet{kraus_2021} using rotating and non-rotating stellar models. Present analysis is limited to the non-rotating evolutionary tracks hence the inclusion of rotation may influence the derived value of the age and the structure of the star. 

Number of unstable modes and pulsation periods associated with these modes are generally affected by the parameters such as mass, surface temperature, luminosity and chemical composition. For linear stability analysis, additional models of MWC 137 have been included to consider the errors present in the determination of surface temperature and luminosity of this star (Fig. \ref{modal_4} and Fig. \ref{modal_5}). 
Variation in the period of the unstable mode as a function of surface temperature for models with 45 M$_{\odot}$ is represented by a secondary horizontal axis created using light brown shaded region in Fig. \ref{fig:evol}. Similarly changes in the period associated with the most unstable mode as a function of luminosity is shown by a secondary vertical axis through a shaded light green region in Fig. \ref{fig:evol}. We infer from Fig. \ref{fig:evol} that for a given mass, models with higher luminosity and lower surface temperature are having longer periods (see also Fig. \ref{modal_4} and Fig. \ref{modal_5}).

\section{Discussion} \label{DC}
In the present study, we have analyzed the TESS light curve of MWC\,137 observed during sector 6 and 33 using  the package `Lightkurve' \citep{2018ascl.soft12013L}. In agreement with \citet{kraus_2021}, we have also found the dominant period of 1.9 d in the data of both sectors for MWC 137.    
To understand the 1.9 d variability in this star, we have constructed three different set of models in the mass range between 27 to 70 M$_{\odot}$ and performed linear stability analysis. Linear stability analysis has revealed that the low order radial modes are unstable. In model-sequence having metallicity Z = 0.02 and Z = 0.03, all the models of MWC\,137 are unstable while in case of Z = 0.01, models having mass in the range of 42 M$_{\odot}$ to 58 M$_{\odot}$ are stable. The number of unstable modes in models of Z = 0.01 is smaller compared to Z = 0.02 and Z = 0.03 which is in accordance with earlier studies where higher metallicity leads to more unstable modes in models of massive stars \citep{kiriakidis1993stability,shiode_2012}. For the case of solar chemical composition, the unstable mode having period of 1.9 d is also present in all the models having mass of 45 M$_{\odot}$ and surface temperature in the range of 4.40 $<$ log T$_{\rm{eff}}$ $<$ 4.49 (Fig. \ref{modal_4}). Therefore the instability present in these models is not being influenced by the  ambiguity present in the measurement of surface temperature for this star. 
Models of MWC 137 having higher luminosity tend to have more unstable modes and the strength of instabilities associated with different unstable modes are increasing with luminosity (Fig. \ref{modal_5}).

Mode interaction phenomena are also evident in all three modal diagrams of the considered models. Generally mode interaction is frequently present in models having high luminosity to mass ratios (more than 10$^{3}$ in solar unit) while the mode interaction in the real part of eigenfrequencies is rare for models of less luminous stars such as $\delta$ scuti stars \citep{yadav_2019}. For the case of $\delta$ scuti stars, real part of the eigenfrequency associated with different consecutive modes are approximately equispaced \citep[see e.g.][]{yadav_2019} while the models of Wolf-Rayet stars, luminous blue variables and other massive stars show frequent mode interaction in their modal diagrams \citep[see e.g.][]{kiriakidis_1996, glatzel_1993c}. Strange mode instabilities \citep{GLATZEL_1998,saio_1998} are often present in mode interaction scenarios. Since unstable modes in the considered models of MWC 137 are also exhibiting mode interaction phenomena therefore we suspect that the found instabilities may be associated with strange modes. However non-adiabatic-reversible (NAR) approximation \citep{gautschy_1990b} will be required to confirm the presence of strange mode instabilities which is beyond the scope of present study. 

 In the linear stability analysis, models of 45M$_{\odot}$ with solar chemical composition has an unstable mode with a period of 1.9 d. This theoretically obtained period is in agreement with the dominant variability as found in TESS data for the star MWC\,137 in the present study and by \citet{kraus_2021}. To determine the final fate of unstable models, we have performed non-linear numerical simulation for a model of this star having mass 45 M$_{\odot}$ and solar chemical composition. In the non-linear regime, the instabilities lead to finite amplitude pulsation with a period of 1.9 d for the considered model (see, Fig. \ref{45m_nonlin}).
 During the evolution of instabilities into the non-linear regime, error in the energy balance should be smaller than the time integrated kinetic and acoustic energies \citep[see e.g.][]{grott_2005,glatzel_2016}.
 Non-linear numerical simulations have been carried out carefully where error in the energy balance is four order of magnitude smaller than the time integrated acoustic energy. Therefore the obtained finite amplitude pulsation with a period of 1.9 d in the model of MWC 137 is resulting from physical instabilities. 
 In addition to the dominant 1.9 d period, MWC 137 also has other variabilities with periods in the range of 3 to 8 d \citep{kraus_2021} likely to be associated with non-radial modes. The variations of quantities mentioned in Fig. \ref{45m_nonlin} and Fig. \ref{45mphot} are resulting from the radial instabilities therefore the direct comparison of variation profiles with observations needs to be done cautiously.

Evolutionary state of MWC\,137 is uncertain. In earlier studies, this star is classified as a pre-main sequence star \citep[see e.g.][]{Hillenbrand_1992,berrilli1992infrared} while in recent studies this star is classified as a post main sequence star \citep[see e.g.][]{mehner_2016,kraus_2017,kraus_2021}. With the help of evolutionary tracks, we find that this star is either a post main sequence star or about to start post main sequence evolution. Evolutionary track which is passing through the average value of surface temperature and luminosity in HR diagram is indicating that the star is burning hydrogen in a shell around helium core. Evolutionary tracks are indicating  that the mass of MWC\,137 is more than 40 M$_{\odot}$ which is in agreement with the model's mass of 45 M$_{\odot}$ showing the pulsation of 1.9 d in linear stability analysis and non-linear numerical simulation. Suggested mass range of 10 to 15 M$_{\odot}$ by \citet{mehner_2016} appears to be quite low while the mass of 45 M$_{\odot}$ is within the range as proposed by \citet{kraus_2021}.

Using the travel distance of 1.2 pc and jet velocity of 650 km/s, \citet{mehner_2016} has estimated that the emanated jet from the central star is at least 1800 yr old. Similarly the authors have also derived the minimum age of 20$\,$000 yr for the large nebula surrounding the central star MWC\,137. The value of log  T$_{\rm{eff}}$ for MWC\,137 ranges between 4.40 to 4.49 \citep{Hillenbrand_1992,Esteban_1988,Alonso-Albi_2009}. Considered stellar evolutionary models having surface temperature in this range are also going through the phase of overall contraction where radius and surface temperature decreases and increases, respectively. Overall contraction phase for the model having mass 48.2  M$_{\odot}$ is of the order of 50\,000 yrs which is longer than the minimum age of the jet and nebula of the central star. Formation of disk around stars and emergence of jets are complex processes which are poorly understood. Generally jet is found in the astrophysical systems having accretion,  magnetic field and rotation \citep{bogovalov_1999,Coffey_2008}. 
The location of overall contraction phase is also partially coinciding with the position of MWC\,137 in the HR diagram. By comparing the involved timescales of overall contraction phase, minimum age of nebula and jet, we speculate that the formation of disk and jets may be linked with the overall contraction phase in this star. To establish this link, further theoretical and observational studies are required. 

\section{Conclusions} \label{conclu}
TESS light curve analysis followed by linear stability analysis and non-linear numerical simulations have been performed for the star MWC 137. The outcome of the present study is summarized in the following four points:

\begin{enumerate}
    \item Similar to the \citet{kraus_2021}, we also find a dominant variability with a period of 1.9 d in TESS data of MWC\,137. This period is present in both  sector 6 and sector 33 of TESS in which this star is observed. 
    
    \item Linear stability analyses performed in models of MWC\,137 show that the low order radial modes are excited  as found in case of other B-type stars \citep{yadav_2017b}. Frequent mode interactions in the modal diagram is indicating the presence of strange mode instabilities in models of this star.
    
    \item Linear stability analyses followed by nonlinear numerical simulations show that a model with mass of 45 M$_{\odot}$ has an excited mode with a period of 1.9 d. Therefore observed dominant variability can be explained using low order radial pulsation mode.  
    
    \item Evolutionary tracks passing through the location of MWC\,137 in the HR diagram is indicating that this star is likely in post main sequence phase or about to end the main sequence evolutionary phase.

\end{enumerate}

\section*{Acknowledgements}
We sincerely acknowledge the referee's valuable comments to improve this paper.
SP, APY, YCJ and SJ gratefully acknowledge the financial support from Core Research Grant (CRG/2021/007772) of Science and Engineering Research Board (SERB), India. MK acknowledges financial support from the Czech Science Foundation (GA\v{C}R, grant number 20-00150S). The Astronomical Institute
of the Czech Academy of Sciences is supported by the project RVO:67985815.
Parts of this research was funded by the European Union's Framework
Programme for Research and Innovation Horizon 2020
(2014-2020) under the Marie Sk\l{}odowska-Curie Grant Agreement No.
823734. 

The used TESS photometric data of MWC 137 are taken from the Mikulski Archive for Space Telescopes (MAST) located at the Space Telescope Science Institute (STScI).

\section*{Data Availability}
Data will be shared on the reasonable request to the authors.



\bibliographystyle{mnras}
\bibliography{example} % if your bibtex file is called example.bib

\begin{thebibliography}{}
\makeatletter
\relax
\def\mn@urlcharsother{\let\do\@makeother \do\$\do\&\do\#\do\^\do\_\do\%\do\~}
\def\mn@doi{\begingroup\mn@urlcharsother \@ifnextchar [ {\mn@doi@} {\mn@doi@[]}}
\def\mn@doi@[#1]#2{\def\@tempa{#1}\ifx\@tempa\@empty \href {http://dx.doi.org/#2} {doi:#2}\else \href {http://dx.doi.org/#2} {#1}\fi \endgroup}
\def\mn@eprint#1#2{\mn@eprint@#1:#2::\@nil}
\def\mn@eprint@arXiv#1{\href {http://arxiv.org/abs/#1} {{\tt arXiv:#1}}}
\def\mn@eprint@dblp#1{\href {http://dblp.uni-trier.de/rec/bibtex/#1.xml} {dblp:#1}}
\def\mn@eprint@#1:#2:#3:#4\@nil{\def\@tempa {#1}\def\@tempb {#2}\def\@tempc {#3}\ifx \@tempc \@empty \let \@tempc \@tempb \let \@tempb \@tempa \fi \ifx \@tempb \@empty \def\@tempb {arXiv}\fi \@ifundefined {mn@eprint@\@tempb}{\@tempb:\@tempc}{\expandafter \expandafter \csname mn@eprint@\@tempb\endcsname \expandafter{\@tempc}}}

\bibitem[\protect\citeauthoryear{{Alonso-Albi, T.}, {Fuente, A.}, {Bachiller, R.}, {Neri, R.}, {Planesas, P.}, {Testi, L.}, {Bern\'e, O.}  \& {Joblin, C.}}{{Alonso-Albi, T.} et~al.}{2009}]{Alonso-Albi_2009}
{Alonso-Albi, T.} {Fuente, A.} {Bachiller, R.} {Neri, R.} {Planesas, P.} {Testi, L.} {Bern\'e, O.}  {Joblin, C.} 2009, \mn@doi [A\&A] {10.1051/0004-6361/200810401}, 497, 117

\bibitem[\protect\citeauthoryear{{Bergner}, {Miroshnichenko}, {Yudin}  \& {Mukanov}}{{Bergner} et~al.}{1994}]{bergner_1994}
{Bergner} Y.~K.,  {Miroshnichenko} A.~S.,  {Yudin} R.~V.,   {Mukanov} D.~B.,  1994, Odessa Astronomical Publications, \href {https://ui.adsabs.harvard.edu/abs/1994OAP.....7...60B} {7, 60}

\bibitem[\protect\citeauthoryear{Berrilli, Corciulo, Ingrosso, Lorenzetti, Nisini  \& Strafella}{Berrilli et~al.}{1992}]{berrilli1992infrared}
Berrilli F.,  Corciulo G.,  Ingrosso G.,  Lorenzetti D.,  Nisini B.,   Strafella F.,  1992, The Astrophysical Journal, 398, 254

\bibitem[\protect\citeauthoryear{Bogovalov \& Tsinganos}{Bogovalov \& Tsinganos}{1999}]{bogovalov_1999}
Bogovalov S.,  Tsinganos K.,  1999, \mn@doi [\mnras] {10.1046/j.1365-8711.1999.02413.x}, 305, 211

\bibitem[\protect\citeauthoryear{{B{\"o}hm-Vitense}}{{B{\"o}hm-Vitense}}{1958}]{bohm_1958}
{B{\"o}hm-Vitense} E.,  1958, \zap, \href {http://adsabs.harvard.edu/abs/1958ZA.....46..108B} {46, 108}

\bibitem[\protect\citeauthoryear{{Ciatti} \& {Mammano}}{{Ciatti} \& {Mammano}}{1975}]{ciatti_1975}
{Ciatti} F.,  {Mammano} A.,  1975, \aap, \href {https://ui.adsabs.harvard.edu/abs/1975A&A....38..435C} {38, 435}

\bibitem[\protect\citeauthoryear{Coffey, Bacciotti  \& Podio}{Coffey et~al.}{2008}]{Coffey_2008}
Coffey D.,  Bacciotti F.,   Podio L.,  2008, \mn@doi [The Astrophysical Journal] {10.1086/592343}, 689, 1112

\bibitem[\protect\citeauthoryear{{Esteban} \& {Fernandez}}{{Esteban} \& {Fernandez}}{1998}]{esteban_1998}
{Esteban} C.,  {Fernandez} M.,  1998, \mn@doi [\mnras] {10.1046/j.1365-8711.1998.01610.x}, \href {https://ui.adsabs.harvard.edu/abs/1998MNRAS.298..185E} {298, 185}

\bibitem[\protect\citeauthoryear{Esteban \& Fernández}{Esteban \& Fernández}{1998}]{Esteban_1988}
Esteban C.,  Fernández M.,  1998, \mn@doi [Monthly Notices of the Royal Astronomical Society] {10.1046/j.1365-8711.1998.01610.x}, 298, 185

\bibitem[\protect\citeauthoryear{{Frost}}{{Frost}}{1902}]{Frost_1902}
{Frost} E.~B.,  1902, \mn@doi [\apj] {10.1086/140929}, \href {https://ui.adsabs.harvard.edu/abs/1902ApJ....15..340F} {15, 340}

\bibitem[\protect\citeauthoryear{{Fuente}, {Rodr{\'\i}guez-Franco}, {Testi}, {Natta}, {Bachiller}  \& {Neri}}{{Fuente} et~al.}{2003}]{fuente_2003}
{Fuente} A.,  {Rodr{\'\i}guez-Franco} A.,  {Testi} L.,  {Natta} A.,  {Bachiller} R.,   {Neri} R.,  2003, \mn@doi [\apjl] {10.1086/380296}, \href {https://ui.adsabs.harvard.edu/abs/2003ApJ...598L..39F} {598, L39}

\bibitem[\protect\citeauthoryear{{Gautschy} \& {Glatzel}}{{Gautschy} \& {Glatzel}}{1990a}]{gautschy_1990a}
{Gautschy} A.,  {Glatzel} W.,  1990a, \mnras, \href {http://adsabs.harvard.edu/abs/1990MNRAS.245..154G} {245, 154}

\bibitem[\protect\citeauthoryear{{Gautschy} \& {Glatzel}}{{Gautschy} \& {Glatzel}}{1990b}]{gautschy_1990b}
{Gautschy} A.,  {Glatzel} W.,  1990b, \mnras, \href {http://adsabs.harvard.edu/abs/1990MNRAS.245..597G} {245, 597}

\bibitem[\protect\citeauthoryear{{Gautschy}, {Glatzel}, {Gautschy}  \& {Glatzel}}{{Gautschy} et~al.}{1990}]{glatzel_1990NAR}
{Gautschy} A.,  {Glatzel} W.,  {Gautschy} A.,   {Glatzel} W.,  1990, \mn@doi [\mnras] {10.1093/mnras/245.4.597}, \href {https://ui.adsabs.harvard.edu/abs/1990MNRAS.245..597G} {245, 597}

\bibitem[\protect\citeauthoryear{{Glatzel}}{{Glatzel}}{1994}]{glatzel_1994}
{Glatzel} W.,  1994, \mn@doi [\mnras] {10.1093/mnras/271.1.66}, \href {https://ui.adsabs.harvard.edu/abs/1994MNRAS.271...66G} {271, 66}

\bibitem[\protect\citeauthoryear{{Glatzel}}{{Glatzel}}{1998}]{GLATZEL_1998}
{Glatzel} W.,  1998, in {Bradley} P.~A.,  {Guzik} J.~A.,  eds,  Astronomical Society of the Pacific Conference Series Vol. 135, A Half Century of Stellar Pulsation Interpretation. p.~89

\bibitem[\protect\citeauthoryear{{Glatzel} \& {Chernigovski}}{{Glatzel} \& {Chernigovski}}{2016}]{glatzel_2016}
{Glatzel} W.,  {Chernigovski} S.,  2016, \mn@doi [\mnras] {10.1093/mnras/stw003}, \href {http://adsabs.harvard.edu/abs/2016MNRAS.457.1190G} {457, 1190}

\bibitem[\protect\citeauthoryear{{Glatzel} \& {Kiriakidis}}{{Glatzel} \& {Kiriakidis}}{1993}]{glatzel_1993c}
{Glatzel} W.,  {Kiriakidis} M.,  1993, \mn@doi [\mnras] {10.1093/mnras/263.2.375}, \href {http://adsabs.harvard.edu/abs/1993MNRAS.263..375G} {263, 375}

\bibitem[\protect\citeauthoryear{{Glatzel}, {Kiriakidis}, {Chernigovskij}  \& {Fricke}}{{Glatzel} et~al.}{1999}]{glatzel_1999}
{Glatzel} W.,  {Kiriakidis} M.,  {Chernigovskij} S.,   {Fricke} K.~J.,  1999, \mn@doi [\mnras] {10.1046/j.1365-8711.1999.02190.x}, \href {http://adsabs.harvard.edu/abs/1999MNRAS.303..116G} {303, 116}

\bibitem[\protect\citeauthoryear{{Goswami} et~al.,}{{Goswami} et~al.}{2022}]{goswami_2022}
{Goswami} S.,  et~al., 2022, \mn@doi [\aap] {10.1051/0004-6361/202142031}, \href {https://ui.adsabs.harvard.edu/abs/2022A&A...663A...1G} {663, A1}

\bibitem[\protect\citeauthoryear{{Grott}, {Chernigovski}  \& {Glatzel}}{{Grott} et~al.}{2005}]{grott_2005}
{Grott} M.,  {Chernigovski} S.,   {Glatzel} W.,  2005, \mn@doi [\mnras] {10.1111/j.1365-2966.2005.09162.x}, \href {http://adsabs.harvard.edu/abs/2005MNRAS.360.1532G} {360, 1532}

\bibitem[\protect\citeauthoryear{Guthnick}{Guthnick}{1913}]{guthnick_1913}
Guthnick P.,  1913, \mn@doi [Astronomische Nachrichten] {https://doi.org/10.1002/asna.19131951404}, 195, 265

\bibitem[\protect\citeauthoryear{{Henning}, {Burkert}, {Launhardt}, {Leinert}  \& {Stecklum}}{{Henning} et~al.}{1998}]{henning_1998}
{Henning} T.,  {Burkert} A.,  {Launhardt} R.,  {Leinert} C.,   {Stecklum} B.,  1998, \aap, \href {https://ui.adsabs.harvard.edu/abs/1998A&A...336..565H} {336, 565}

\bibitem[\protect\citeauthoryear{{Hillenbrand}, {Strom}, {Vrba}  \& {Keene}}{{Hillenbrand} et~al.}{1992}]{Hillenbrand_1992}
{Hillenbrand} L.~A.,  {Strom} S.~E.,  {Vrba} F.~J.,   {Keene} J.,  1992, \mn@doi [\apj] {10.1086/171819}, \href {https://ui.adsabs.harvard.edu/abs/1992ApJ...397..613H} {397, 613}

\bibitem[\protect\citeauthoryear{{Iglesias} \& {Rogers}}{{Iglesias} \& {Rogers}}{1996}]{iglesias_1996}
{Iglesias} C.~A.,  {Rogers} F.~J.,  1996, \mn@doi [\apj] {10.1086/177381}, \href {http://adsabs.harvard.edu/abs/1996ApJ...464..943I} {464, 943}

\bibitem[\protect\citeauthoryear{Kiriakidis, Fricke  \& Glatzel}{Kiriakidis et~al.}{1993}]{kiriakidis1993stability}
Kiriakidis M.,  Fricke K.,   Glatzel W.,  1993, Monthly Notices of the Royal Astronomical Society, 264, 50

\bibitem[\protect\citeauthoryear{{Kiriakidis}, {Glatzel}  \& {Fricke}}{{Kiriakidis} et~al.}{1996}]{kiriakidis_1996}
{Kiriakidis} M.,  {Glatzel} W.,   {Fricke} K.~J.,  1996, \mn@doi [\mnras] {10.1093/mnras/281.2.406}, \href {http://cdsads.u-strasbg.fr/abs/1996MNRAS.281..406K} {281, 406}

\bibitem[\protect\citeauthoryear{{Kraus}}{{Kraus}}{2009}]{kraus_2009}
{Kraus} M.,  2009, \mn@doi [\aap] {10.1051/0004-6361:200811020}, \href {https://ui.adsabs.harvard.edu/abs/2009A&A...494..253K} {494, 253}

\bibitem[\protect\citeauthoryear{{Kraus}}{{Kraus}}{2019}]{kraus_2019}
{Kraus} M.,  2019, \mn@doi [Galax] {10.3390/galaxies7040083}, \href {https://ui.adsabs.harvard.edu/abs/2019Galax...7...83K} {7, 83}

\bibitem[\protect\citeauthoryear{{Kraus} et~al.,}{{Kraus} et~al.}{2017}]{kraus_2017}
{Kraus} M.,  et~al., 2017, \mn@doi [\aj] {10.3847/1538-3881/aa8df6}, \href {https://ui.adsabs.harvard.edu/abs/2017AJ....154..186K} {154, 186}

\bibitem[\protect\citeauthoryear{{Kraus}, {Liimets}, {Moiseev}, {S{\'a}nchez Arias}, {Nickeler}, {Cidale}  \& {Jones}}{{Kraus} et~al.}{2021}]{kraus_2021}
{Kraus} M.,  {Liimets} T.,  {Moiseev} A.,  {S{\'a}nchez Arias} J.~P.,  {Nickeler} D.~H.,  {Cidale} L.~S.,   {Jones} D.,  2021, \mn@doi [\aj] {10.3847/1538-3881/ac1355}, \href {https://ui.adsabs.harvard.edu/abs/2021AJ....162..150K} {162, 150}

\bibitem[\protect\citeauthoryear{{Lenz} \& {Breger}}{{Lenz} \& {Breger}}{2005}]{periodo4}
{Lenz} P.,  {Breger} M.,  2005, \mn@doi [Communications in Asteroseismology] {10.1553/cia146s53}, \href {https://ui.adsabs.harvard.edu/abs/2005CoAst.146...53L} {146, 53}

\bibitem[\protect\citeauthoryear{{Liermann}, {Kraus}, {Schnurr}  \& {Fernandes}}{{Liermann} et~al.}{2010}]{liermann_2010}
{Liermann} A.,  {Kraus} M.,  {Schnurr} O.,   {Fernandes} M.~B.,  2010, \mn@doi [\mnras] {10.1111/j.1745-3933.2010.00915.x}, \href {https://ui.adsabs.harvard.edu/abs/2010MNRAS.408L...6L} {408, L6}

\bibitem[\protect\citeauthoryear{{Lightkurve Collaboration} et~al.,}{{Lightkurve Collaboration} et~al.}{2018}]{2018ascl.soft12013L}
{Lightkurve Collaboration} et~al., 2018, {Lightkurve: Kepler and TESS time series analysis in Python}, Astrophysics Source Code Library (\mn@eprint {ascl} {1812.013})

\bibitem[\protect\citeauthoryear{{Marston} \& {McCollum}}{{Marston} \& {McCollum}}{2006}]{marston_2006}
{Marston} A.~P.,  {McCollum} B.,  2006, in {Kraus} M.,  {Miroshnichenko} A.~S.,  eds,  Astronomical Society of the Pacific Conference Series Vol. 355, Stars with the B[e] Phenomenon. p.~189

\bibitem[\protect\citeauthoryear{{Marston} \& {McCollum}}{{Marston} \& {McCollum}}{2008}]{marston_2008}
{Marston} A.~P.,  {McCollum} B.,  2008, \mn@doi [\aap] {10.1051/0004-6361:20066086}, \href {https://ui.adsabs.harvard.edu/abs/2008A&A...477..193M} {477, 193}

\bibitem[\protect\citeauthoryear{{Mehner} et~al.,}{{Mehner} et~al.}{2016}]{mehner_2016}
{Mehner} A.,  et~al., 2016, \mn@doi [\aap] {10.1051/0004-6361/201527180}, \href {https://ui.adsabs.harvard.edu/abs/2016A&A...585A..81M} {585, A81}

\bibitem[\protect\citeauthoryear{{Miroshnichenko}}{{Miroshnichenko}}{1994}]{miro_1994}
{Miroshnichenko} S.,  1994, in {The} P.~S.,  {Perez} M.~R.,   {van den Heuvel} E. P.~J.,  eds,  Astronomical Society of the Pacific Conference Series Vol. 62, The Nature and Evolutionary Status of Herbig Ae/Be Stars. p.~134

\bibitem[\protect\citeauthoryear{{Muratore}, {Kraus}, {Liermann}, {Schnurr}, {Cidale}  \& {Arias}}{{Muratore} et~al.}{2010}]{muratore_2010a}
{Muratore} M.~F.,  {Kraus} M.,  {Liermann} A.,  {Schnurr} O.,  {Cidale} L.~S.,   {Arias} M.~L.,  2010, Boletin de la Asociacion Argentina de Astronomia La Plata Argentina, \href {https://ui.adsabs.harvard.edu/abs/2010BAAA...53..123M} {53, 123}

\bibitem[\protect\citeauthoryear{{Muratore}, {Kraus}, {Oksala}, {Arias}, {Cidale}, {Borges Fernandes}  \& {Liermann}}{{Muratore} et~al.}{2015}]{muratore_2015}
{Muratore} M.~F.,  {Kraus} M.,  {Oksala} M.~E.,  {Arias} M.~L.,  {Cidale} L.,  {Borges Fernandes} M.,   {Liermann} A.,  2015, \mn@doi [\aj] {10.1088/0004-6256/149/1/13}, \href {https://ui.adsabs.harvard.edu/abs/2015AJ....149...13M} {149, 13}

\bibitem[\protect\citeauthoryear{{Nomoto}, {Kobayashi}  \& {Tominaga}}{{Nomoto} et~al.}{2013}]{nomoto_2013}
{Nomoto} K.,  {Kobayashi} C.,   {Tominaga} N.,  2013, \mn@doi [\araa] {10.1146/annurev-astro-082812-140956}, \href {http://adsabs.harvard.edu/abs/2013ARA%26A..51..457N} {51, 457}

\bibitem[\protect\citeauthoryear{{Oksala}, {Kraus}, {Cidale}, {Muratore}  \& {Borges Fernandes}}{{Oksala} et~al.}{2013}]{oksala_2013}
{Oksala} M.~E.,  {Kraus} M.,  {Cidale} L.~S.,  {Muratore} M.~F.,   {Borges Fernandes} M.,  2013, \mn@doi [\aap] {10.1051/0004-6361/201321568}, \href {https://ui.adsabs.harvard.edu/abs/2013A&A...558A..17O} {558, A17}

\bibitem[\protect\citeauthoryear{{Paxton}}{{Paxton}}{2004}]{paxton_2004}
{Paxton} B.,  2004, \mn@doi [\pasp] {10.1086/422345}, \href {https://ui.adsabs.harvard.edu/abs/2004PASP..116..699P} {116, 699}

\bibitem[\protect\citeauthoryear{{Ricker} et~al.,}{{Ricker} et~al.}{2015}]{Ricker_2015}
{Ricker} G.~R.,  et~al., 2015, \mn@doi [Journal of Astronomical Telescopes, Instruments, and Systems] {10.1117/1.JATIS.1.1.014003}, \href {https://ui.adsabs.harvard.edu/abs/2015JATIS...1a4003R} {1, 014003}

\bibitem[\protect\citeauthoryear{{Rogers} \& {Iglesias}}{{Rogers} \& {Iglesias}}{1992}]{rogers_1992}
{Rogers} F.~J.,  {Iglesias} C.~A.,  1992, \mn@doi [\apjs] {10.1086/191659}, \href {http://adsabs.harvard.edu/abs/1992ApJS...79..507R} {79, 507}

\bibitem[\protect\citeauthoryear{{Rogers}, {Swenson}  \& {Iglesias}}{{Rogers} et~al.}{1996}]{rogers_1996}
{Rogers} F.~J.,  {Swenson} F.~J.,   {Iglesias} C.~A.,  1996, \mn@doi [\apj] {10.1086/176705}, \href {http://adsabs.harvard.edu/abs/1996ApJ...456..902R} {456, 902}

\bibitem[\protect\citeauthoryear{{Saio}, {Baker}  \& {Gautschy}}{{Saio} et~al.}{1998}]{saio_1998}
{Saio} H.,  {Baker} N.~H.,   {Gautschy} A.,  1998, \mn@doi [\mnras] {10.1111/j.1365-8711.1998.01195.x}, \href {https://ui.adsabs.harvard.edu/abs/1998MNRAS.294..622S} {294, 622}

\bibitem[\protect\citeauthoryear{{Sharpless}}{{Sharpless}}{1959}]{sharpless_1959}
{Sharpless} S.,  1959, \mn@doi [\apjs] {10.1086/190049}, \href {https://ui.adsabs.harvard.edu/abs/1959ApJS....4..257S} {4, 257}

\bibitem[\protect\citeauthoryear{Shiode, Quataert  \& Arras}{Shiode et~al.}{2012}]{shiode_2012}
Shiode J.~H.,  Quataert E.,   Arras P.,  2012, \mn@doi [Monthly Notices of the Royal Astronomical Society] {10.1111/j.1365-2966.2012.21130.x}, 423, 3397

\bibitem[\protect\citeauthoryear{{Skinner}, {Brown}  \& {Stewart}}{{Skinner} et~al.}{1993}]{skinner_1993}
{Skinner} S.~L.,  {Brown} A.,   {Stewart} R.~T.,  1993, \mn@doi [\apjs] {10.1086/191803}, \href {https://ui.adsabs.harvard.edu/abs/1993ApJS...87..217S} {87, 217}

\bibitem[\protect\citeauthoryear{{Th\'e}, {de Winter}  \& {Perez}}{{Th\'e} et~al.}{1994}]{the_1994}
{Th\'e} P.~S.,  {de Winter} D.,   {Perez} M.~R.,  1994, \aaps, \href {https://ui.adsabs.harvard.edu/abs/1994A&AS..104..315T} {104, 315}

\bibitem[\protect\citeauthoryear{{Ulrich} \& {Wood}}{{Ulrich} \& {Wood}}{1981}]{ulrich_1981}
{Ulrich} R.~K.,  {Wood} B.~C.,  1981, \mn@doi [\apj] {10.1086/158692}, \href {https://ui.adsabs.harvard.edu/abs/1981ApJ...244..147U} {244, 147}

\bibitem[\protect\citeauthoryear{{Yadav}}{{Yadav}}{2019}]{yadav_2019}
{Yadav} A.~P.,  2019, Bulletin de la Societe Royale des Sciences de Liege, \href {https://ui.adsabs.harvard.edu/abs/2019BSRSL..88..110Y} {88, 110}

\bibitem[\protect\citeauthoryear{{Yadav} \& {Glatzel}}{{Yadav} \& {Glatzel}}{2016}]{yadav_2016}
{Yadav} A.~P.,  {Glatzel} W.,  2016, \mn@doi [\mnras] {10.1093/mnras/stw236}, \href {http://adsabs.harvard.edu/abs/2016MNRAS.457.4330Y} {457, 4330}

\bibitem[\protect\citeauthoryear{{Yadav} \& {Glatzel}}{{Yadav} \& {Glatzel}}{2017}]{yadav_2017b}
{Yadav} A.~P.,  {Glatzel} W.,  2017, \mn@doi [\mnras] {10.1093/mnras/stx1808}, \href {http://adsabs.harvard.edu/abs/2017MNRAS.471.3245Y} {471, 3245}

\bibitem[\protect\citeauthoryear{{Yadav}, {K{\"u}hnrich Biavatti}  \& {Glatzel}}{{Yadav} et~al.}{2018}]{yadav_2018}
{Yadav} A.~P.,  {K{\"u}hnrich Biavatti} S.~H.,   {Glatzel} W.,  2018, \mn@doi [\mnras] {10.1093/mnras/sty092}, \href {http://cdsads.u-strasbg.fr/abs/2018MNRAS.475.4881Y} {475, 4881}

\bibitem[\protect\citeauthoryear{{Yadav}, {Joshi}  \& {Glatzel}}{{Yadav} et~al.}{2021}]{yadav_2021}
{Yadav} A.~P.,  {Joshi} S.,   {Glatzel} W.,  2021, \mn@doi [\mnras] {10.1093/mnras/staa3611}, \href {https://ui.adsabs.harvard.edu/abs/2021MNRAS.500.5515Y} {500, 5515}

\makeatother
\end{thebibliography}

%\begin{thebibliography}{99}
%\end{thebibliography}

%%%%%%%%%%%%%%%%%%%%%%%%%%%%%%%%%%%%%%%%%%%%%%%%%%

%%%%%%%%%%%%%%%%% APPENDICES %%%%%%%%%%%%%%%%%%%%%

\appendix

%%%%%%%%%%%%%%%%%%%%%%%%%%%%%%%%%%%%%%%%%%%%%%%%%%

% Don't change these lines
\bsp	% typesetting comment
\label{lastpage}
\end{document}